\documentstyle[psfig,multicol,aps]{revtex}
\begin{document}
%\draft
\title{Pairing and the structure of the $pf$-shell $N\sim Z$ nuclei}
\author{Alfredo Poves$^{\dagger}$ and Gabriel
  Martinez-Pinedo$^{\ddagger}$\cite{dire}}
\address{$^{\dagger}$Departamento de F\'{\i}sica Te\'orica, \\
  Universidad Aut\'onoma, Cantoblanco, 28049 Madrid, Spain}

\address{$^{\ddagger}$ W.K. Kellogg Radiation Laboratory, 106-38,
  California Institute of Technology \\
  Pasadena, California 91125 USA}

\date{\today}
\maketitle

\begin{abstract}
  The influence of the isoscalar and isovector $L=0$ pairing
  components of the effective nucleon-nucleon interaction is evaluated
  for several isobaric chains, in the framework of full $pf$ shell
  model calculations.  We show that the combined effect of both
  isospin channels of the pairing force is responsible for the
  appearance of $T=1$ ground states in $N=Z$ odd-odd nuclei. However,
  no evidence is found relating them to the Wigner energy.  We study
  the dependence of their contributions to the total energy on the
  rotational frecuency in the deformed nucleus $^{48}$Cr. Both
  decrease with increasing angular momentum and go to zero at the band
  termination.  Below the backbending their net effect is a reduction
  of the moment of inertia, more than half of which comes from the
  proton-neutron channel.
\end{abstract}
\pacs{PACS number(s): }

\begin{multicols}{2}

The study of the isovector pairing interaction among like particles
is one of the classical themes of nuclear physics. Proton-neutron
pairing, has been much less studied, in particular its isoscalar part.
Early attempts to solve the isoscalar and isovector $L=0$ pairing in
one $l$ shell can be found in ref. \cite{evans1} and an extension to
the two shells case in \cite{evans2}.  The experimental access to
medium mass $N=Z$ nuclei, approaching the proton drip line has renewed
the interest for the components of the nuclear interaction whose
manifestations are enhanced close to $N=Z$. The proton-neutron part of
the isovector pairing near $N=Z$ has been studied in term of algebraic
model and compared with Shell Model Monte Carlo calculations in
ref.~\cite{langanke}.

In particular there are many discussions on the role of the isoscalar
pairing interaction.  Is it responsible for the Wigner energy? Should
it give rise to a proton-neutron condensate? How will its
manifestations evolve as the angular momentum increases?. Some of
these issues have been examined in refs.~\cite{satu,wyss}. The role of
the isoscalar pairing channel in the processes mediated by the
Gamow-Teller operator has been recently studied in ref. \cite{engel}.

In this letter we take advantage of the availability of detailed
microscopic descriptions of the nuclei in the middle of the $pf$-shell
-that have already passed many experimental tests- to address some of
these questions \cite{sm1,sm2}. The existence of well deformed rotors
in the $A=50$ mass region makes it possible to explore the dependence
of a given observable with the rotational
frequency~\cite{sm6,sm7,lenzi,silvia}.

We shall study the isobaric chains $A=46$, 48 and 50, using the
spherical shell model approach. The details of the shell model work
have been explained in ref. \cite{sm1}. We use the Lanczos shell model
code ANTOINE (ref. \cite{antoine}) and the the effective interaction
KB3. All the calculations presented are unrestricted in the
$pf$-shell.
 
A key ingredient of our approach is the choice of the ``physical''
isocalar and isovector pairing terms of the effective nuclear
interaction. We follow here the work of Dufour and Zuker in
ref~\cite{duzu}, where they give a precise characterization of the
multipole part of the nuclear interaction, showing in particular that
the pairing term of any realistic G-matrix has only two contributions;
the well known schematic isovector $L=0$, $S=0$ pairing and an
isoscalar part which turns out to be again the schematic $L=0$, $S=1$
pairing.  This eliminates many ambiguities associated to the choice of
the isovector and isoscalar pairing interactions. We shall keep the
notation in~\cite{duzu} and use P01 for the isovector and P10 for the
isoscalar pairing. The explicit expressions for the two body matrix
elements in LS and $jj$ couplings are:

\begin{mathletters}
\begin{eqnarray}
  \label{eq:p01LS}
  &\langle& l_A l_B L S J T|\text{P01}| l_C l_D L S J T\rangle
  \nonumber \\
  & & = G \sqrt{(2 l_A +1) (2 l_C +1)} \delta_{AB} \delta_{CD}
  \delta_{L0} \delta_{S0} \delta_{J0} \delta_{T1},
\end{eqnarray}  

\begin{eqnarray}
  \label{eq:p10LS}
  &\langle& l_A l_B L S J T|\text{P10}| l_C l_D L S J T\rangle
  \nonumber \\
  & & = G \sqrt{(2 l_A +1) (2 l_D +1)} \delta_{AB} \delta_{CD}
  \delta_{L0} \delta_{S1} \delta_{J1} \delta_{T0},
\end{eqnarray}  

\begin{eqnarray}
  \label{eq:p01jj}
  &\langle& j_a j_b J T|\text{P01}| j_c j_d J T\rangle \nonumber \\
  & &= G \sqrt{(j_a +1/2) (j_c +1/2)} \delta_{ab} \delta_{cd}
  \delta_{J0} \delta_{T1},
\end{eqnarray}  

\begin{eqnarray}
  \label{eq:p10jj}
  &\langle& j_a j_b J T|\text{P10}| j_c j_d J T\rangle\nonumber\\
  & & = G \frac{2
    (-1)^{j_a-j_c}}{\sqrt{1+\delta_{ab}}\sqrt{1+\delta_{cd}}}
  \sqrt{(2j_a+1) (2j_b+1) (2j_c+1) (2j_d+1)} \nonumber \\[2mm] 
   & & \times \left\{\begin{array}{ccc}
      1/2 & j_a & l_a\\
      j_b & 1/2 & 1
    \end{array}\right\}
  \left\{\begin{array}{ccc}
      1/2 & j_c & l_c\\
      j_d & 1/2 & 1 
    \end{array}\right\} \delta_{AB} \delta_{CD} \delta_{J1}
      \delta_{T0}, 
\end{eqnarray}  
\end{mathletters}
where capital letters denote a $l$-orbit (quantum numbers $nl$) and
lowercase letters denote a $j$-orbit (quantum numbers $nlj$).  The
value for the pairing constant $G$ is easily obtained from the numbers
given in table I of the same reference. We use $G=-0.295$ for P01 and
$G=-0.459$ for P10. It could be thinkable to do the same kind of
analysis with a monopole free pairing interaction. We have verified
that none of the conclusions of the paper would be modified if we did
so. Thus, we decided to stick to the usual forms of pairing.
 
Once this definition of the pairing operators adopted, we focuss in
the role they play in the behaviour of different nuclei and different
physical quantities. For that we shall proceed in the most
straightforward way; for each case we make first a reference
calculation using the interaction KB3, then we substract from KB3 the
isovector or the isoscalar pairing hamiltonians and make the
calculations with the new interactions KB3--P01 and KB3--P10. We
obtain the effect of each pairing channel by direct comparison with
the reference calculation. One could also estimate these effects in
perturbation theory, in the cases studied here we have verified that
the results are equivalent, however this may not be true in general.

\vskip 1cm
\noindent
{\bf Binding energies for an isobaric chain.}

\begin{figure}
  \begin{center}
    \leavevmode
    \psfig{file=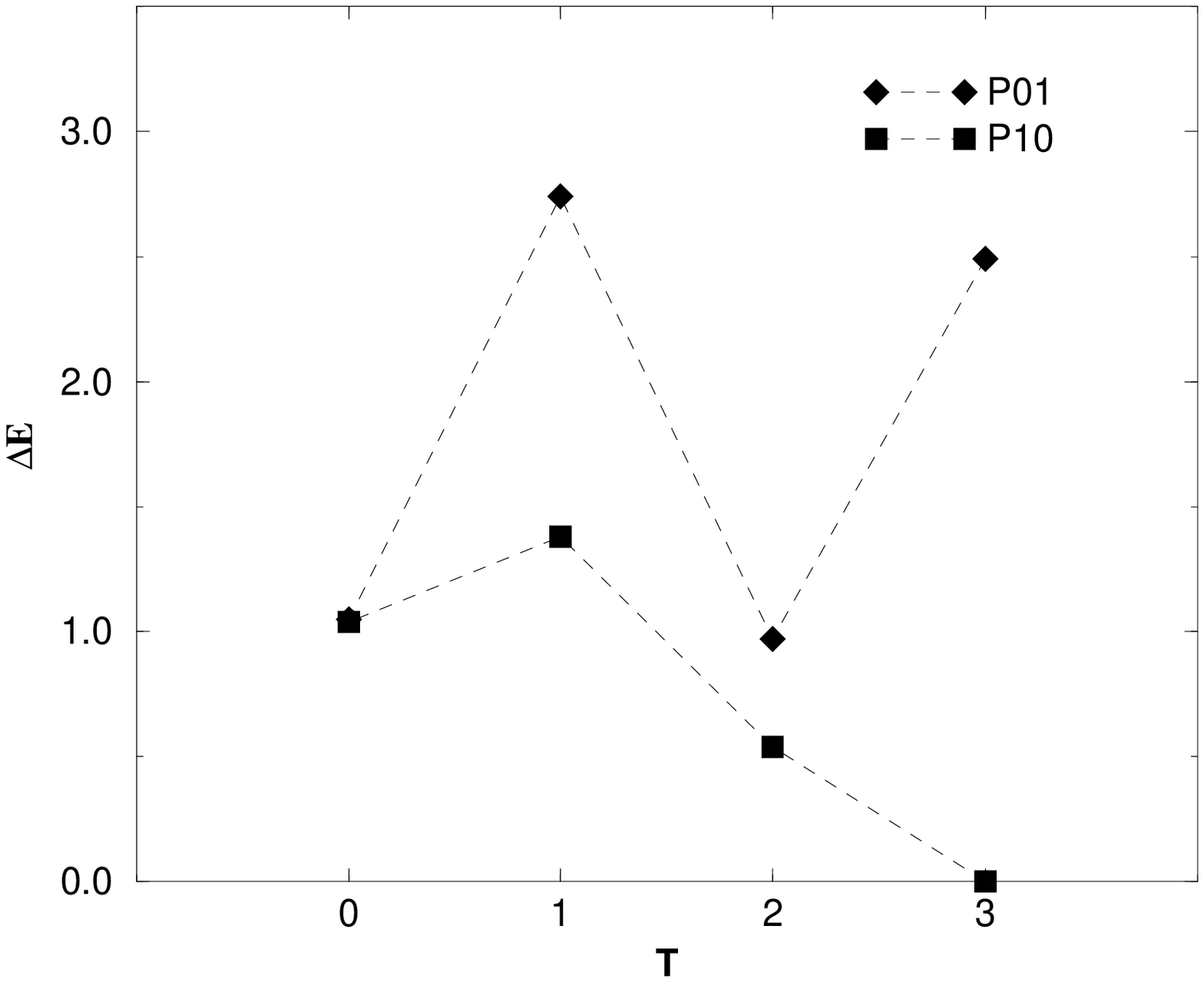,width=0.45\textwidth}
    \caption{Pairing contributions to the ground state energies of
      the A=46 isobaric multiplet (in MeV)}
    \label{fig:a46}
  \end{center}
\end{figure}

We begin with $A=46$ in figure~\ref{fig:a46}. We have plotted the
contributions to the ground state binding energy (BE) from the
isovector pairing (labeled P01) and from the isoscalar pairing
(labeled P10). BE's are taken with the positive sign. The figure shows
the strong odd-even staggering of the P01 points, mostly suppressed in
the P10 case, as expected. The effect of P10, although important, is
smaller than that of P01 and goes smoothly to zero as the number of
valence protons goes to zero (notice that our description is fully
simetric under the exchange of protons and neutrons). The only little
surprise is that moving from $T=1$ to $T=0$, not only the contribution
of P01 decreases, but also the contibution of P10, although to a
lesser extent. This may explain why the ground state of $^{46}$V has
$T=1$ instead of the expected $T=0$. It goes like this; the monopole
part of KB3 will put the centroid of the $T=0$ states lower than the
centroid of the $T=1$ states by about 1.3 MeV; on the other hand, the
total pairing contribution to $T=1$ is larger by more than 1.5~MeV
than the contribution to $T=0$ (see figure~\ref{fig:a46}), therefore
it is finally the $T=1$ state that becomes the ground state of the
odd-odd $T_z=0$ member of the multiplet. Notice that the inversion
depends on the balance between the isovector monopole term and both
pairings. As the isovector monopole has a smooth dependence with mass,
while the isovector pairing depends linearly with the degeneracy of
the orbit being filled, we expect the inversion to take place when
high $j$ orbits are dominant. When low $j$ orbits are the relevant
ones, the isoscalar pairing contribution to the $T=0$ state can be
enhanced due to the small spin-orbit splitting, what helps producing
$T=0$ ground states. This is borne out by the experiments; when the
$1f_{7/2}$ orbit is being filled the $T=1$ states are ground states,
in $^{58}$Cu the $2p_{3/2}$ orbit is dominant and restores a $T=0$
ground state, in $^{62}$Ga the influence of the $1f_{5/2}$ orbit
shifts again to a $T=1$ ground state. From there on the $1g_{9/2}$
orbit starts to play an important role and $T=1$ ground states occur
in all cases.  Actually the $T=0$-$T=1$ splitting could give a hint on
the ordering of the single particle orbits in places where no other
information is available.

\begin{figure}
    \begin{center}
      \leavevmode
      \psfig{file=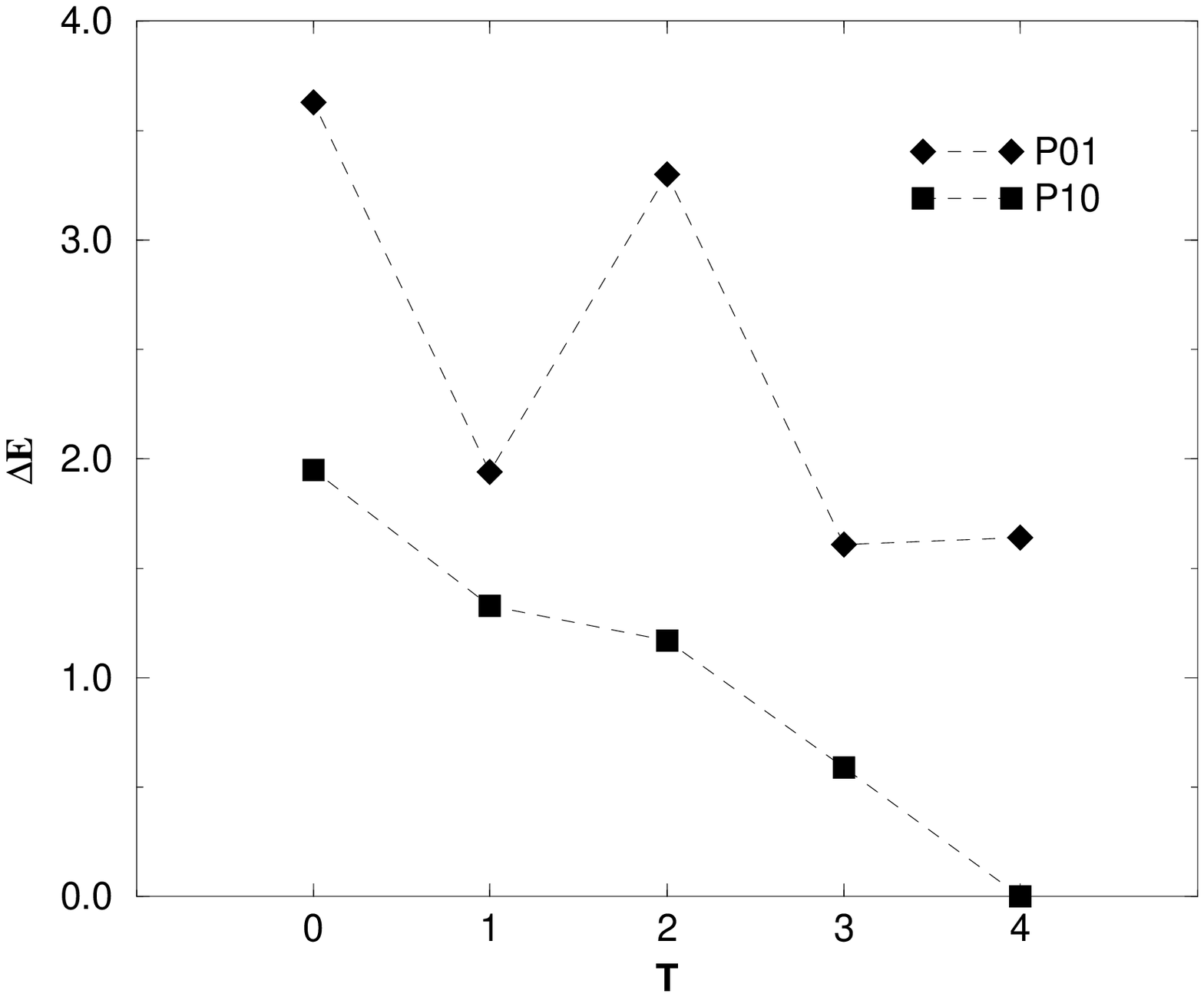,width=0.45\textwidth}
      \caption{Pairing contributions to the ground state energies of
       the A=48 isobaric multiplet (in MeV)}
      \label{fig:a48}
    \end{center}
\end{figure}

\begin{figure}
    \begin{center}
      \leavevmode
      \psfig{file=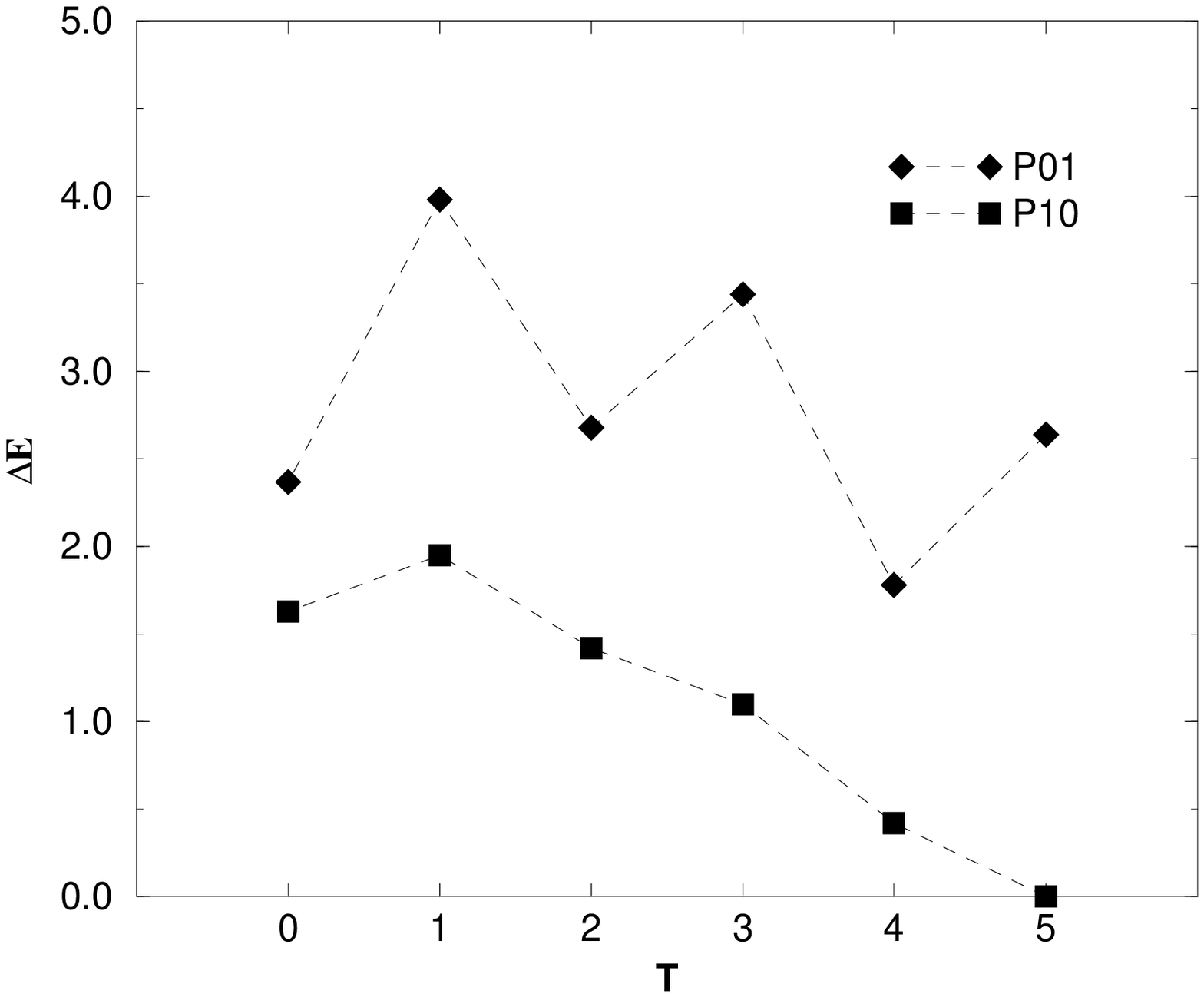,width=0.45\textwidth}
      \caption{Pairing contributions to the ground state energies of
      the A=50 isobaric multiplet (in MeV)}
      \label{fig:a50}
    \end{center}
\end{figure}

In figure~\ref{fig:a48} the same quantities are plotted for $A=48$.
The behaviour of the P10 contribution is very smooth and we do not
find any breaking of slope at $T=0$. The P01 points stagger as usual.
In figure~\ref{fig:a50} we present $A=50$. Now the part of the
isobaric chain lying inside the $pf$-shell is longer and the trends
seen in $A=46$ better established. There is a slight increase of the
P10 contribution from $T=0$ to $T=1$ followed by a smooth decrease,
reaching zero at $T=5$. As in $A=46$ the reduction of both P10 and P01
contributions from $T=1$ to $T=0$ explains the occurrence of a $T=1$
ground state in $^{50}$Mn.

\vskip 1cm
\noindent 
{\bf Wigner energy.}

This name is given to the extra binding of the $T=0$ member of an
isobaric multiplet compared to what would be predicted by an $(N-Z)^2$
extrapolation from the other isobars. We can make an assesment of the
role of the P01 and P10 terms proceeding as in the definition just
given. In $A=46$ we have too few nuclei in the valence space as to
make the fit, therefore we only present results for the two other
chains.

\noindent
 {\bf For $A=48$},

\begin{quote}
$\epsilon_W$ = 2.82 \hspace{1cm}  KB3

$\epsilon_W$ = 3.04 \hspace{1cm}  KB3--P01

$\epsilon_W$ = 2.47 \hspace{1cm}  KB3--P10
\end{quote}

\noindent
clearly the effect of both types of pairing on the Wigner energy is
negligible.

\noindent
 {\bf For $A=50$},

\begin{quote}
$\epsilon_W$ = 1.99 \hspace{1cm}  KB3

$\epsilon_W$ = 2.60 \hspace{1cm}  KB3--P01

$\epsilon_W$ = 2.12 \hspace{1cm}  KB3--P10
\end{quote}

\noindent
the effect of the pairing terms is even less significant than in the
$A=48$ isobars. In conclusion, we find no link between the Wigner
energy and the dominant pairing terms of the nuclear interaction.
Furthermore, if we had made the fit with the more microscopic
prescription $T(T+1)$ the values of this ``new'' Wigner energy would
have been much smaller (1.23 and 0.24 for $A=48$ and $A=50$) and the
effect of the pairing terms negligible again. We therefore think that
the Wigner energy cannot be explained as a pairing effect.

\vskip 1cm
\noindent
{\bf Pairing and angular momentum}
 
\begin{figure}
    \begin{center}
      \leavevmode
      \psfig{file=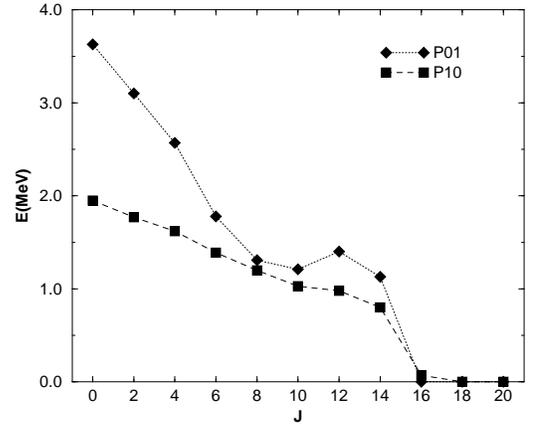,width=0.45\textwidth}
      \caption{Pairing contributions to the energies of the yrast
      states in $^{48}$Cr.}
      \label{fig:cr48_1}
    \end{center}
\end{figure}

\begin{figure}
    \begin{center}
      \leavevmode
      \psfig{file=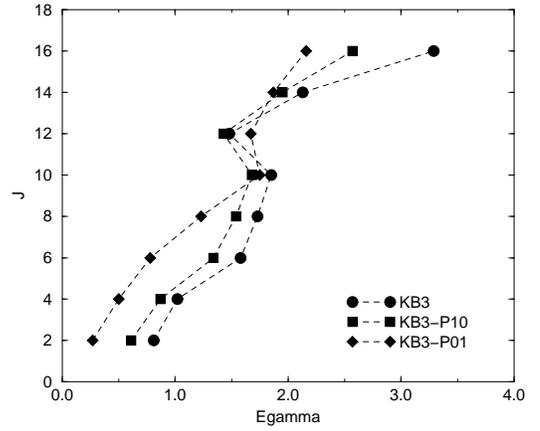,width=0.45\textwidth}
      \caption{Gamma ray energies along the yrast band of $^{48}$Cr
      (in MeV);
      full interaction (KB3); without isoscalar pairing (KB3-P10);
      without isovector pairing (KB3-P01)}
      \label{fig:cr48_2}
    \end{center}
\end{figure}

The evolution of the pair content of a nucleus as the rotational
frequency increases has been the subject of many heated debates in the
recent past. We have studied this evolution in several cases but we
shall present only results for $^{48}$Cr, which is the most
representative example. In figure~\ref{fig:cr48_1} we plot the
contributions of the P10 and P01 pairing hamiltonians to the absolute
energy of the yrast levels. Two regions show up, corresponding roughly
to below/above the backbending. In the region of collective rotation
the P01 contribution, that at $J=0$ almost doubles the P10, decreases
with $J$ much more rapidly than the P10 one. At $J=8$ both reach
approximately the same value, and stay constant until they suddenly
vanish at the aligned $J=16$ state. In figure~\ref{fig:cr48_2} we have
translated the results of figure~\ref{fig:cr48_1} to a backbending
plot. The most visible effect of the pairing correlation is the well
known reduction of the moment of inertia. The effect of the P01
channel is clearly dominant but the P10 contribution is far from being
negligible and amounts to about one fourth of the P01.  This suggest a
plausible explanation to the discrepancies found in the comparison of
the Cranked Hartree Fock Bogolyubov results using the Gogny force and
the Shell Model (or the experimental ones) for the yrst band of
$^{48}$Cr \cite{sm6}. Both approaches were shown to predict correctly
the quadrupole properties, however the CHFB calculation gave a much
larger moment of inertia.  Actually it came out a factor three larger
than the experiment, which is fully compatible with what we have seen
to be the effect of taking out the proton neutron pairing ($T=0$ and
$T=1$).  Indeed, proton-neutron pairing is not properly taken into
account in the standard HFB treatements.  $^{50}$Cr and $^{50}$Mn
behave in a qualitatively similar way. The only difference being the
smaller role played by the P10 pairing in the moment of inertia of
$^{50}$Cr.

\vskip 1cm
\noindent
{\bf Pairing condensates?}
 
There has been speculations recently on the possible existence of a
new superfluid phase of nuclear matter made of a condensate of $L=0$,
$S=1$ neutron proton pairs, that could be approachable in $N=Z$
nuclei. Once the meaning of ``condensate'' agreed, we can try to
contribute to the debate.

Imagine that the $1f_{7/2}$ orbit were the valence space and P01 the
effective interaction, in this case, the ground state of an even-even
nucleus will be the seniority zero state that can be easily
interpreted as a condensate of $L=0$, $S=0$, $T=1$ pairs (with all the
caveats, because only the valence particles are paired and the $j$'s
of the orbits are not large enough compared with the number of active
particles).  Given that the $(1f_{7/2})^n$ configurations are the
leading ones in the ground states of the nuclei we are interested in,
it is in principle possible that the ``P01 condensate''. i.e. the
ground state of the $(1f_{7/2})^n$ isovector pairing problem, and the
physical ground state could have a reasonable overlap. Furthermore,
most of the properties of the $(1f_{7/2})^n$ pairing condensate
persist if we solve the pairing problem in the full $pf$-shell with
the experimental single particle splitting.

On the contrary, if we solve the hamiltonian P10 in the $(1f_{7/2})^n$
configurations, we do not find any condensate. In order to obtain a
condensate of $L=0$, $S=1$, $T=0$ pairs, the valence space must
include at least the $1f_{5/2}$ orbit degenerate with the $1f_{7/2}$,
or, in general to be made of $l$-orbits, which is an extremely
unphysical situation.  Taking this argument to its limit one could say
that the existence of a triplet pairing condensate is excluded by the
spin orbit splitting of the nuclear mean field.

We have tried to make these statements quantitative computing the
overlaps of the wave fuctions of the two condensates and the physical
ground state in several $pf$-shell nuclei.  Let's call
$|\text{KB3}\rangle$ the physical ground state of a given nucleus
obtained with the interaction KB3; $| \text{CP01}\rangle$ the
isovector condensate obtained solving P01 in the $pf$ shell with the
experimental single particle splitting and $|\text{CP10}\rangle$ the
isoscalar condensate obtained solving P10 in the $pf$ shell without
spin-orbit splitting, i.e. $l=3$ and $l=1$ separated by 2.0~MeV.  The
squared overlaps of these states with the physical ground state give
the probabilities of finding the nuclei in the different condensed
phases.  The results are gathered in table I and the conclusions
rather evident. The isovector $L=0$ pairing condensate clearly
dominates the structure of the Calcium isotopes we have selected. When
two protons are added the overlaps get reduced and the interpretation
in term of condensates is less justified.  For the Chromiums no trace
of a condensate is left. Notice that an overlap of 38$\%$ can be
considered almost marginal. Actually, the overlap of the $^{48}$Cr
ground state, obtained using a pure $Q \cdot Q$ interaction and the
isovector pairing condensate is already 25$\%$. As anticipated the L=0
isoscalar pairing condensate does not show up, on the contrary its is
almost exactly orthogonal to the physical ground states in most of the
cases.  As we said before, the existence of such a condensate would
only be possible if in the valence space that accomodates the physics
of the problem, the spin-orbit splittings were negligible.

\vskip 1cm
\noindent
{\bf Conclusions}

We have explored the effect of the isovector and isoscalar $L=0$
pairing interactions on several properties of $pf$-shell nuclei close
to $N=Z$.  We have found that the isoscalar pairing, whose
contribution to the ground state binding energy of the nuclei studied
may be very important, is not at the origin of the Wigner energy. Nor
the isovector term. We argue that the occurrence of $T=1$ ground
states in odd-odd, $N=Z$ nuclei is due to the combined action of the
two pairing channels that, when the dominant orbit has large $l$, can
overcome the isovector monopole gap. We stress the need of a good
treatment of the proton-neutron pairing in order to obtain the correct
moment of inertia of $N=Z$ rotors. Finally we discard the presence of
isoscalar pairing condensates everywhere in the region, as well as the
existence of isovector pairing condensates at $N=Z$.

\acknowledgements 

The authors thank P. Vogel, K.  Langanke, S.  Koonin, E. Caurier and
A. Zuker for many interesting discussions.  G.M. thanks the Spanish
Ministerio de Educacion y Ciencia for a postdoctoral fellowship. This
work has been partly supported by a grant of DGES (Spain) PB96-053.

\narrowtext

\begin{table}
  \begin{center}
    \caption{Overlaps between the physical ground state and the singlet
      and triplet pairing ``condensates'' in several nuclei.}
    \label{tab:over}
    \begin{tabular}{cccc}
      nucleus &  $|\langle \text{KB3}|\text{CP01} \rangle|^2$
      &  $|\langle \text{KB3}|\text{CP10} \rangle|^2$
      &  $|\langle \text{CP10}|\text{CP01} \rangle|^2$ \\
      \hline
      $^{44}$Ca & 97  &  -  &  -  \\
      $^{46}$Ca & 95  &  -  &  -  \\
      $^{44}$Ti & 63  & 16  &  1 \\
      $^{46}$Ti & 52  & 10  &  1 \\
      $^{48}$Ti & 59  &  1  &  0  \\
      $^{48}$Cr & 38  & 2   &  0 \\
      $^{50}$Cr & 42  & 1   &  0 
    \end{tabular}
  \end{center}
\end{table}

\end{multicols}

\end{document}